\documentclass[twocolumn]{emulateapj}
\usepackage{graphicx}
\usepackage{epsfig}
\usepackage{amsmath}
\usepackage{natbib}
\usepackage[dvips]{color}

\newcommand{\msun}{M_{\odot}}

\begin{document}
\title{Tidal Disruption of Protoclusters in Giant Molecular Clouds}
\author{Eva-Marie Proszkow}
\affil{Physics Department, University of Michigan, Ann Arbor, MI; emdavid@umich.edu}
\and
\author{Philip C. Myers}
\affil{Harvard-Smithsonian Center for Astrophysics, Cambridge, MA}

\shorttitle{Tidal Disruption of Protoclusters}
\shortauthors{Proszkow \& Myers}

\begin{abstract}

We study the collapse of protoclusters within a giant molecular cloud (GMC) to determine the conditions under which collapse is significantly disrupted.  Motivated by observations of star forming regions which exhibit flattened cloud structures, this study considers collapsing protoclusters with disk geometries. The collapse of a $10^3 \msun$ protocluster initially a distance of $2-10$ pc from a $10^3 - 10^6 \msun$ point mass is numerically calculated.  Simulations with zero initial relative velocity between the two are completed as well as simulations with relative velocities consistent with those observed in GMCs.  The results allow us to define the conditions under which it is safe to assume protocluster collapse proceeds as if in isolation. For instance, we find the collapse of a $10^3 \msun$ protocluster will be significantly disrupted if it is within $2-4$ pc of a $10^4 \msun$ point mass.  Thus, the collapse of a $10^3 \msun$ protocluster can be considered to proceed as if in isolation if it is more than $\sim 4$ pc away from a $10^4 \msun$ compact object.  In addition, in no portion of the sampled parameter space does the gravitational interaction between the protocluster disk and the massive particle significantly disperse the disk into the background GMC. We discuss the distribution of clusters of young stellar objects within the Perseus and Mon R2 star forming regions, which are consistent with the results of our simulations and the limitations of our results in gas dominated regions such as the Orion cloud.

\end{abstract}

\section{Introduction} \label{sec:Introduction}

Over the last three decades it has become evident that molecular clouds serve as the site of all star formation within galaxies with the result that a clear understanding of molecular cloud structure and evolution is necessary to fully describe the initial conditions for star formation.  Typical giant molecular clouds (GMCs) are $\sim 10 - 50$ pc across and have masses between $10^6$ and $10^7 \msun$\citep{Liszt1981ApJ,Solomon1987ApJ}.  They are nonuniform, composed of high density clumps with sizes of  $\sim 1 - 10$ pc and masses of $\sim 10^2 - 10^5 M_{\odot}$ embedded in a lower density background \citep{Kramer1998AA,Heyer1998ApJ}.  These high density clumps are the sites of star formation within GMCs.  

Although GMCs do not globally collapse, many of the small-scale high density clumps undergo local collapse and fragmentation.  The most massive of these clumps produce clusters of young stars \citep{Ballesteros-Paredes2007PPV}.  Throughout this paper, the term ``protocluster'' will refer to these massive, high density regions within molecular clouds which have the potential to produce clusters of stars.  The process by which protoclusters fragment into collapsing substructures and eventually protostars is complex and not completely understood.  It may be due to a combination of Jeans instabilities, the decay of turbulence \citep{Klessen2000ApJS, Klessen2001ApJ}, and the decoupling of fluid and MHD waves \citep{Myers1998ApJL}.  Observations of star forming regions reveal that newly formed clusters of stars are not isolated within molecular clouds.  Rather, star forming regions of molecular clouds tend to contain multiple clusters and smaller groups of young stellar objects near each other \citep{Allen2007PPV}.

Since GMCs are highly nonuniform in structure and there is evidence of clusters forming near other clusters, a relevant question in cluster formation is: under what conditions can the collapse of a protocluster be treated in isolation, i.e. ignoring the influence of the surrounding GMC environment? 

This question is analogous to the question of how star and planet formation in cluster environments differs from formation in isolation.  Studies that have considered star and planet formation within clustered environments have shown that environment may have modest to significant effects on the formation processes depending on the properties of cluster considered.  Specifically, in very dense massive clusters frequent close encounters between stars may disrupt disks and young solar systems and UV radiation from massive stars may cause photoevaporation of protostellar disks \citep{Bonnell1998APh, Storzer1999ApJ}.  However, in more intermediate-sized clusters, with parameters typical of those observed within $\sim 1$ kpc of the sun,  interactions between stars are much less common, and the cluster's effects on planet-forming disks are relatively modest \citep{Adams2006ApJ}.  In this study, we consider a similar question, but on a larger scale and at an earlier time in cluster development.

The above question can be addressed in a gross way with simple tidal force estimates, i.e. calculating the Roche limit.  However the analytic Roche limit does not provide the details of how tidal distortion affects the evolution of the protocluster structure, how this evolution varies with the initial cloud structure, nor how it depends on the relative motion of the protocluster with respect to its environment.  To address these questions, one must complete numerical simulations of the collapsing protocluster within its GMC environment.

To this end, we perform an ensemble of $N$-body simulations to calculate the collapse of a protocluster in the presence of a nearby massive object in the same GMC. The point mass is intended to represent another protocluster or high density region that is sufficiently massive and compact enough to be roughly modeled as a point mass. These calculations treat only the gravitational interactions and ignore the effects of gas pressure, stellar winds, and radiation.  Molecular cloud maps of star forming regions contain numerous dense gas structures with varying geometries, so a full numerical investigation of this question should consider protoclusters with spherical, flattened, and filamentary initial geometries.  In this study, we focus on flattened disk-shaped protoclusters.  

Our choice of protocluster geometry is partially motivated by observations of star forming regions revealing flattened, layered, and/or filamentary cloud structures.  For example shells of molecular gas surrounding OB stars have been identified by \cite{Deharveng2005AA} and \cite{Churchwell2006ApJ}. Within many of these shells triggered star formation appears to be occurring \citep{Churchwell2006ApJ, Zavagno2006AA}.  Furthermore, the distribution of young stars within embedded clusters is often aligned with the elongation or filament structure of the embedding gas \citep{Allen2007PPV, Kumar2007MNRAS, Teixeira2006ApJL}. Observations such as these suggest that ``reduced dimensionality'' may be a characteristic of star forming clouds. 

Flattened cloud structures may be formed by OB winds \citep{Weaver1977ApJ, Whitworth1994AA} or expanding H \textsc{ii} regions \citep{Elmegreen1977ApJ} sweeping up and condensing nearby cold molecular gas.  Cluster forming clouds that are believed to have been compressed by these mechanisms include Orion B near the OB1 association \citep{Wilson2005AA} and the DR 21 ridge near the Cyg OB2 association \citep{Schneider2006AA, Kumar2007MNRAS}.  \cite{Elmegreen1998Proceed} and \cite{Whitworth2005Proceed} provide more detailed discussions of proposed flattening mechanisms capable of producing the structure observed in many star forming regions.  Simulations by \cite{Burkert2004ApJ} demonstrated that even in isolation, finite self-gravitating sheets collapse into dense structures with interesting geometries.

In addition to the observational evidence for flattened initial states, we choose to simulate flattened protoclusters because they are harder to disrupt than spherical protoclusters of the same mass and radius.  This is because the disks are more centrally concentrated and thus more tightly bound.  Therefore they are more likely to succeed in forming stellar clusters than their 3D counterparts, all else being equal. 

We simulate the collapse of a disk-shaped protocluster in the presence of another dense object in the GMC to determine the survivability of collapsing protoclusters in GMCs.   In $\S$ \ref{sec:SimDescription} the $N$-body collapse calculations are described.   The results from these simulations are summarized in $\S$ \ref{sec:Results}, and a discussion of these results in the context of observed star forming regions is presented in $\S$ \ref{sec:Discussion}.  The Appendix contains a discussion of the simulations of collapsing systems with analytic solutions used to estimate the uncertainties in $N$-body calculations of the dynamics of gaseous systems.


\section{Numerical Calculation of Disk Collapse} \label{sec:SimDescription}

In this study, we complete an ensemble of $N$-body simulations to study the collapse of flattened protoclusters in the presence of a nearby dense object represented by a massive point particle.  This dense object may be another dense protocluster, a region that is currently forming stars, a stellar cluster, or another significant density enhancement in the GMC, provided that it is more massive than the protocluster and small enough that its gravitational influence on the protocluster can be well approximated as a point mass.   A modified version of the NBODY2 direct $N$-body integration code \citep{Aarseth2001NewAstro, Aarseth2003Book} is used to complete the simulations. The NBody2 code was modified to implement a sink cell algorithm associated with the particle initially located at the center of the protocluster disk.  This algorithm allows particles that collapse to the center of the disk to accrete onto it. A more detailed discussion of the sink cell algorithm in included in the Appendix.

A fluid dynamics code will always model a fluid system with more precision than will an $N$-body code because on small scales close encounters between point masses deviate from fluid behavior.  For this reason $N$-body techniques are not widely used to study fluid problems.  Two significant concerns arise when modeling a fluid system with an $N$-body code.  First, $N$-body simulations do not take into account the pressure forces between point particles.  Secondly, close gravitational scattering encounters between point particles are not physically realistic in fluid systems.  If these two concerns are appropriately addressed, an $N$-body code may be competitive with a full fluid simulation when studying the large scale gravitational evolution of a fluid.

Specifically, if the fluid can be approximated as cold and pressure free, the dominant force between fluid elements is gravity and thus the fluid elements may be modeled as individual particles moving under the influence of their mutual gravitational potential.  Protoclusters are cold with typical temperatures of $\lesssim 20$ K \citep{Liszt1981ApJ, Solomon1987ApJ} and can be approximated as pressure free fluids until the late stages of collapse.  Our choice of radially symmetric ring structure was chosen specifically to minimize the number of close scattering interactions that particles undergo during disk evolution (For a more detailed discussion, see Appendix.)  Thus we can carefully use an $N$-body code to address  the large scale question of when the collapse of a protocluster may be treated as if it occurs in isolation, neglecting the effects of the surrounding GMC environment.  

In addition, for modest values of $N$ (in this study, $N \lesssim 1000$) $N$-body simulations are computationally more efficient than fluid codes. Therefore, judicious use of an $N$-body code with a relatively modest investment of computational time can return a broad understanding of a system over a wide range of parameter space.

To better understand the regime in which the $N$-body simulation reliably models gaseous disk collapse behavior and thus can be used effectively in this study, the code must be systematically tested.  Therefore, this particular $N$-body code was used to complete simulations of collapsing systems for which analytic solutions exist.  Specifically,  the collapse of an isolated Maclaurin disk and Mestel disk were simulated and the results of the simulations compared to the analytic collapse solutions.  We find that the $N$-body simulations predict free fall times for the Maclaurin and Mestel disks with errors less than $3.4\%$ and $2.4\%$ and cumulative mass profiles with errors less than $6.1\%$ and $7.4\%$, respectively.   A complete discussion of these test calculations and comparison to the pressure-free analytic collapse solutions is provided in the Appendix and summarized in Table \ref{tab:Errors}.  

It is important to stress that this approach investigates only the gravitational interactions within the disk and between the disk and the point mass.  Other physical processes in gas such as pressure and turbulence are not included in our simulations of protocluster collapse.  The detailed structure of the final cluster forming cloud and the amount of star formation which will subsequently occur depends on these processes.  A more in depth discussion of how these physical processes may affect the results of our simulations is reserved for $\S$ \ref{sec:DiscussOtherProcess}.

We assume that the flattened protoclusters are formed when a three dimensional dense cloud is compressed into a two dimensional disk.  This assumption is motivated by evidence that flattened structures in star forming regions may be created by stellar winds or shocks sweeping material into a layer. Thus the surface density of the protocluster disk depends on the assumed density of the pre-flattened three dimensional cloud from which it is formed.   Consider a uniform density sphere which is compressed along one dimension. The resulting density distribution, a Maclaurin disk, has the surface density profile
\begin{equation}
\Sigma(r) = \left\{ \begin{array}{ll}
\Sigma(0) \left[ 1- \left(\frac{r}{r_d}\right)^2 \right]^{1/2}, & 0 \leq r \leq r_d \\
0, & r > r_d, \\
\end{array} \right.
\end{equation}
where $r_d$ is the disk radius \citep{BT1987Book}.  

The majority of our simulations considered disks that are initially described by Maclaurin surface density profiles.  These disks are weakly centrally condensed and consistent with the disk formation scenario described above.  It is difficult to form a less centrally condensed disk distribution by flattening a realistic three dimensional clump of gas.  However, flattening of a somewhat centrally condensed three dimensional clump of gas will produce an even more strongly centrally concentrated disk.  To investigate the evolutionary differences central concentration may create, a limited number of strongly centrally concentrated disks were also completed. Specifically we considered the pre-flattened three dimensional dense cloud with the density distribution of a singular isothermal sphere.  When flattened it produces a Mestel disk described by the surface density distribution
\begin{equation}
\Sigma(r) = \frac{\Sigma(A)A}{r}, 0 < r < \infty,
\end{equation}
where $\Sigma(A)$ is the surface density at the fiducial radius $A$ \citep{Mestel1963MNRAS}.  A comparison of the Maclaurin and Mestel disk evolution is included in $\S$ \ref{sec:DiscussCompare}. 

The free fall time for a particle initially at rest at radius $r \leq r_d$ in a Maclaurin disk of mass $M_d$ is 
\begin{equation}
t_{ff} (r) = \sqrt{\frac{\pi}{6}}\sqrt{\frac{r_d^3}{GM_d}}.
\end{equation}
Therefore, a Maclaurin disk collapses all at once.  The free fall time for a particle initially at rest at radius $r$ in a Mestel disk is
\begin{equation}
t_{ff}(r) = \frac{\sqrt{\pi}}{4}\frac{r}{\sqrt{G\Sigma(A)A}}
\end{equation}
resulting in an inside-out collapse.  

Each simulated disk is composed of a central point mass and 25 rings of 32 equal mass particles.  The mass of the central particle and of each ring is varied to produce the desired surface density distribution.  Each disk has a total mass of $M_d = 10^3 \msun$ and a radius of $r_d = 1$ pc. (The simulated Mestel disk is truncated at $r_d$.)  For star formation efficiencies of $\sim 0.3$ these parameters are consistent with observations of young clusters with a stellar masses of a few hundred solar masses and radii on the order of a parsec \citep{Lada2003ARAA, Porras2003AJ}.  For Maclaurin and Mestel disks with these parameters the free fall time of a particle initially at rest at $r_d$ is $0.342$ Myr and $0.525$ Myr respectively.

In this study we investigate the disruption of a protocluster disk due a massive point particle $M_c$ which represents a neighboring high density region in the GMC a distance $d$ from the center of the disk and in the plane defined by the disk.  We choose this initial geometry because the most efficient way to tidally disrupt an axisymmetric disk is by placing the massive perturber in the plane of the disk.  A point mass coincident with the symmetry axis will encourage disk collapse or, if the separation distance is small enough, accretion onto the massive particle.  A point mass that is neither coincident with the symmetry axis nor in the disk plane will tidally distort the disk, but not as strongly as a point mass at the same distance but located in the plane of the disk.

We consider disks that have both zero and nonzero initial velocities with respect to the nearby point mass, point masses within the range $M_c = 1 - 10^3 M_d ~(10^3 - 10^6 \msun)$, and separation distances of $d = 2-10 r_d ~(2- 10$ pc).  The majority of the simulations consider disks with Maclaurin density profiles.  A limited number of simulations were also completed using the more centrally condensed Mestel density distribution. The results for the Maclaurin and Mestel disks are compared in $\S$ \ref{sec:Results}.  Disk collapse begins at time $t = 0$ and continues until the disk is accreted onto the neighboring point mass or reaches a maximum mean surface density.  This maximum value occurs at approximately the disk's free-fall time and is explained in more detail in $\S$ \ref{sec:DenseEnhanc}.

By modeling the nearby high density region as a point mass, we overestimate its effect on the collapsing disk.   The point mass representation is only exact for density enhancements that have spherically symmetric mass distributions (and a handful of special non-spherical mass distributions).  The high density clump in the GMC will likely be a distributed mass which does not fall into one of these special cases, and will exert a force on the disk that is less than the force of a point particle with a mass equal to the mass of the clump and located at the clump's center.  Therefore, our estimates of tidal disruption may not be applicable in regions containing larger amounts of distributed dense gas, for example the Orion (L1641) region (see $\S$ \ref{sec:Discussion} for further discussion).

A large body of work exists concerning the destructive effects galaxy and GMC environment can have on stellar clusters.  \cite{Terlevich1987MNRAS} and \cite{Theuns1991MmSAI} completed $N$-body simulations to investigate the effect the galactic tidal field and interstellar clouds have on the long-term evolution of initially bound stellar clusters.  More recent studies have considered the disruption of a stellar cluster due to interactions with passing spiral arms and GMCs \citep{Gieles2006MNRAS, Gieles2007MNRAS}.  In addition, there have been many numerical studies of the disruption of stellar clusters when the original embedding gas is violently removed by winds from massive stars or nearby supernovae \citep{Lada1984ApJ, Geyer2001MNRAS, Adams2000ApJ}.  These works consider the interaction of stellar clusters and stellar cluster members with their environments while our calculations consider an earlier stage in the life of a stellar cluster.  Specifically we determine the effects of GMC environment on the clouds that are possible progenitors of stellar clusters.


\section{Simulation Results} \label{sec:Results}

The initial simulations consider the behavior of a collapsing Maclaurin disk in the presence of a massive point particle for various mass ratios and separation distances.  For comparison, a limited number of simulations of collapsing Mestel disks were also completed.  The results of these calculations are summarized below.


\subsection{Interaction Outcomes}\label{sec:IntOutcomes}

\begin{figure}
  \plotone{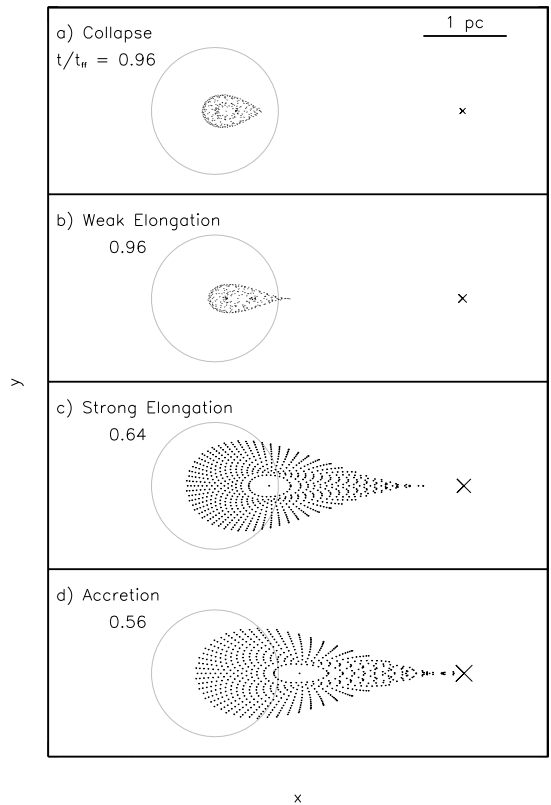}
	\caption{Face-on view of the collapse of a Maclaurin disk $3.0$ pc away from a high mass particle.  Each point denotes the position of a particle within the disk.  Panel a): \emph{Collapse:} undistorted collapse of disk to a point, point mass $M_c = 5 M_d$.  Panel b): \emph{Weak Elongation:} collapse of a disk to short filament, point mass $M_c = 10 M_d$.   Panel c): \emph{Strong Elongation:} stretching of disk into a long filament, point mass $M_c = 50 M_d$.  Panel d): \emph{Accretion:} accretion of disk onto a nearby massive particle, point mass $M_c = 100 M_d$. The location of the point mass is marked by an X where the size of the X indicates the mass.  For each simulation, the display time which most clearly represents the evolution scenario was chosen.  The times are noted in units of the analytic free fall time of the disk in isolation ($t_{ff} = 0.342$ Myr).  The size and position of the disk at time $t = 0$ is indicated by the circle.}
	\label{fig:Snapshots}
\end{figure}

In each simulation, the disk evolves according to one of three distinct behavior patterns: \emph{collapse}, \emph{elongation}, or \emph{accretion}.  The disk evolution depends most sensitively on the mass of the neighboring point mass and the initial distance of separation (see $\S$ \ref{sec:DependMd2} for further discussion.)  Figure \ref{fig:Snapshots} provides a face on view of disks evolving via different scenarios.  For clarity, all of the disks depicted in Figure \ref{fig:Snapshots} are initially $3.0$ pc from a massive particle, and only the mass of the particle is varied.

Figure \ref{fig:Snapshots} (a) depicts the evolution of a disk that follows the \emph{collapse} evolution scenario.  This disk collapses to a point-like object with little to no distortion in the disk's shape.  A small amount of elongation along the $x$-axis does occur.  This is to be expected as it is the axis which intersects the center of the disk and the massive point particle. However, the ratio of the short axis of the disk to the long axis of the disk ($(\Delta y/\Delta x)_{min}$, see $\S$ \ref{sec:Elongation}) remains greater than $0.5$ for all disks that evolve via the collapse scenario.  

\emph{Elongation} refers to disks that collapse while being tidally squeezed. These disks result in filament-like dense structures.  The elongation scenario is further subdivided into two categories: weak and strong elongation.  \emph{Weak elongation} occurs when the disk collapses into a filament that is shorter than the initial diameter of the disk. On the other hand, \emph{strong elongation} occurs when the disk is stretched into a filament that is longer than the initial disk diameter.  The latter behavior indicates that the force of tidal disruption is stronger than the force of collapse.

Figure \ref{fig:Snapshots} (b) displays the evolution of a Maclaurin disk that undergoes weak elongation.  The disk collapses as in the collapse scenario, but at earlier times shows significant distortion.  The differences between strong and weak elongation become apparent when Figure \ref{fig:Snapshots} (b) is compared to Figure \ref{fig:Snapshots} (c).  A disk that is strongly elongated becomes significantly tidally stretched.  Tidal interaction, rather than gravitational collapse dominates the disks evolution.  In addition, a strongly elongated disk accelerates toward the system's center of mass more than a weakly elongated disk, which is another indication of the strength of the force between the disk and the neighboring point mass.

The evolution of a disk \emph{accreting} onto the neighboring point mass is shown in Figure \ref{fig:Snapshots} (d).  The disk is significantly stretched out by the tidal force of the massive neighbor.  Moreover, it travels the initial separation distance quickly enough that the disk does not have time to collapse under its own self-gravity (as in the collapse or weak elongation scenarios), or reach a maximum density as a filament (as in the strong elongation scenario).

\subsection{Dependence of Outcome on Neighbor Mass and Distance}\label{sec:DependMd2}

\begin{figure}
  \plotone{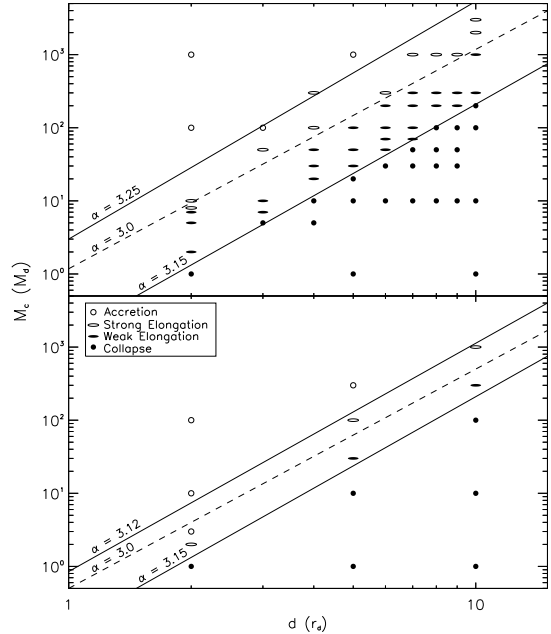}
	\caption{Maclaurin (upper panel) and Mestel (lower panel) disk collapse behavior as a function of the mass of the point mass $M_c$ in units of disk mass and the initial separation $d$ in units of disk radius. Filled circles correspond to disks that collapse to point-like objects with little or no distortion due to the nearby massive point mass, \emph{Collapse}.  Filled ellipses correspond to disks that collapse to filament-like structures under the influence of a nearby massive point mass, \emph{Weak Elongation}.  Open ellipses correspond to disks that are tidally stretched to long filament-like structures under the influence of a nearby point mass, \emph{Strong Elongation}.  Open circles correspond to disks that accrete onto the nearby point mass before significant collapse occurs, \emph{Accretion}. Solid lines separate the portions of the $M$-$d$ space in which systems exhibit significantly different collapse behaviors.  The dashed line in each panel indicates the Roche limit. The slopes $\alpha$ of the solid and dashed lines are indicated in each panel.}
	\label{fig:md2behavior}
\end{figure}

Figure \ref{fig:md2behavior} displays the behavior of the Maclaurin and Mestel disks as a function of the influencing particle mass, $M_c$, and the initial separation distance, $d$.  The portions of $M$-$d$ space in which the systems behave differently are separated by lines in the figure.  Systems that reside in the lower right portion of the plot will have disks that collapse without disruption or distortion from the nearby clump.  This is the region of $M$-$d$ space in which protocluster disk collapse occurs as if in isolation.  The band of $M$-$d$ space stretching from low mass and low separation distance to high mass and high separation distance is the region in which the disk undergoes elongation into a filament as it collapses.  Systems residing within this regime should form filaments stretched out toward the massive nearby high density regions.  Finally, systems with properties in the upper left portion of the $M$-$d$ plot are those in which the protocluster disks will accrete onto the nearby massive object.  We should not expect to find individually collapsing protoclusters near massive dense objects with these properties.

A constant value of initial force between the center of the disk and the massive particle corresponds to a slope of $2$ in Figure \ref{fig:md2behavior} due to the inverse-square nature of the gravitational force.  The dividing lines between regions of different system behavior have slopes slightly steeper than $2$ due to the geometry of the interacting system.  For smaller values of $d$, the disk has less distance to travel before it will accrete onto the nearby point mass.  Furthermore, for smaller values of $d$ the point mass will more effectively squeeze the disk along the $y$-axis and stretch the disk along the $x$-axis (where the axes are defined as in Figure \ref{fig:Snapshots}).  These processes combined cause two systems with equal values of $M_c/d^2$ but different initial separations to behave differently.  Specifically, the system with smaller $d$ will be more easily distorted by the nearby point mass, which is consistent with the results of our simulations.  

Calculation of the Roche limit  also provides insight into the behavior of a collapsing disk under the influence of a nearby massive object.  In our nomenclature, the Roche limit is the distance $d$ at which an object with radius $r_d$ and mass $M_d$ will be torn apart by a point mass $M_c$ \citep{Shu1982Book}.  Specifically, it is the distance at which the tidal force from the nearby point mass is equal to the force that causes the disk to collapse under its own self-gravity.  The calculation of the Roche limit depends on the density distribution of the disk and is given by $d_R = \kappa(M_d/M_c)^{1/3}r_d$ where $\kappa = (8/3\pi)^{1/3}$ for the Maclaurin disk and $\kappa = 2^{1/3}$ for the Mestel disk.  In Figure \ref{fig:md2behavior} the Roche limit is denoted by a dashed line with slope $3$ in each panel.  Note that the Roche limit approximately corresponds to the separation between the weak and strong elongation.  This further supports the distinction between the regime in which the tidal disruption dominates the system behavior and the regime in which self-gravitational collapse dominates the disk behavior.

The evolution of a Mestel and Maclaurin disks are very similar for a given disk-to-point mass ratio and initial separation.  However, the Mestel disks are slightly more easily disrupted.  Specifically, accretion and significant elongation occur at more modest mass ratios for a given initial separation, $d$.  Comparison of the inside-out collapse of a Mestel disk to the uniform collapse of the Maclaurin disk provides insight into why this is so.  The free fall time for the outer portion of a Mestel disk is about $50\%$ longer than that of a corresponding Maclaurin disk.  Therefore the exterior of the Mestel disk is exposed to the tidal forces due to the nearby point mass for a longer period of time.  In addition, the Mestel disk has a longer amount of time to travel toward the massive particle before it collapses, more easily coming near enough for edge accretion to occur.  This difference is also borne out by the Roche limit calculation:  the power law coefficient $\kappa$ is larger for the Mestel disk which indicates that tidal disruption occurs at larger distances for the Mestel disk than for the Maclaurin disk.  It is important to note that although the outer regions of the Mestel disk are more vulnerable to distortion and accretion, the interior of a Mestel disk is extremely robust against tidal distortion.  Even a Mestel disk that is stretched into a filament (strong elongation) contains a massive core which is a remnant of the rapid collapse of the inner portion of the disk.

\subsection{Disk Density Enhancement}\label{sec:DenseEnhanc}

\begin{figure}
  \plotone{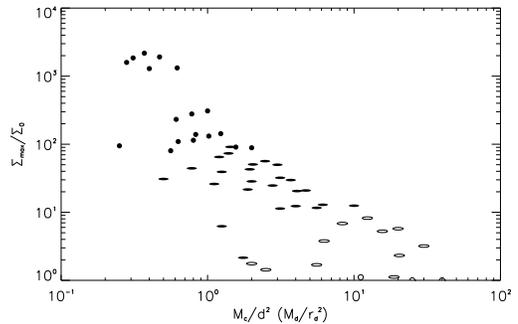}
	\caption{Maximum density enhancement factor as a function of initial disk-point mass configuration in terms of $M_c/d^2$ (in units of $M_d/r_d^2$) for the Maclaurin disk simulations.  Plotting symbols correspond to behavior of disk-point mass system as in Figure \ref{fig:md2behavior}.}
	\label{fig:sigmamax}
\end{figure}

Figure \ref{fig:sigmamax} presents the maximum surface density enhancement factor for each simulated Maclaurin disk over the course of the disk's evolution or distortion.  The density enhancement factor $\Sigma(t)/\Sigma_0$ is defined as the ratio of the mean surface density at time $t$ to the initial mean surface density.  To calculate the mean surface density one of the outer rings in the disk is identified as the boundary of the disk.  The average surface density is then the ratio of the mass contained interior to that ring (including the mass of the ring itself) to the area of the convex hull that encircles all particles interior to the chosen ring.  The convex hull of a set of points is the smallest convex set that contains all of the points in the set, and can be informally thought of as the polygon created by stretching a rubber band around the outside of a set of points \citep{Ripley1977JAP}.   For the data presented in Figure \ref{fig:sigmamax} the boundary chosen to calculate the surface density enhancement factor is the smallest ring containing at least $90\%$ of the mass (We do not choose the ring that encloses $100\%$ of the mass in order to reduce ``edge effects'' discussed in the Appendix).  

As disk collapse proceeds the value of $\Sigma(t)/\Sigma_0$ increases as the particles in the disk approach the disk's center.  An artifact of the simulation technique is that some disk particles pass close to the sink cell without entering it and continue outward on a radial path away from the disk's center.  Thus the mean density reaches a maximum (at approximately the analytic free fall time for the disk collapse in isolation) and then begins to decrease.  This is not commensurate with the behavior of a fluid which continues to collapse until pressure forces slow and eventually halt the collapse.  Therefore, when a collapsing disk reaches a maximum density we consider this to be the end of the collapse phase. 

While the evolution scenarios give general information about the fate and morphology of disrupted protoclusters, the surface density enhancement factor gives more detailed information about the amount of collapse that occurs within the protocluster.  The disks that collapse to a point with little to no distortion attain the highest values of $\Sigma_{max}/\Sigma_0$. The Mestel disks attain values of $\Sigma_{max}/\Sigma_0$ which are significantly higher than the Maclaurin disks.  This again is due to the fact that the Mestel disk collapses from the inside-out.  The collapse produces a massive central core which is not even slightly distorted due to the tidal stretching/squeezing of the nearby massive object.  For larger values of $M_c/d^2$ the maximum density enhancement factor drops due to the more significant interactions between the collapsing disk and the point mass.  The filamentary structures created during these interactions have much lower surface densities than the point-like structures.  Disks that do not collapse onto themselves but instead accrete directly onto the nearby point mass have disk surface densities that decrease as a function of time, and thus attain values of $\Sigma_{max}/\Sigma_0 \sim 1$ near the beginning of the disk's evolution.

\subsection{Disk Elongation}\label{sec:Elongation}

\begin{figure}
  \plotone{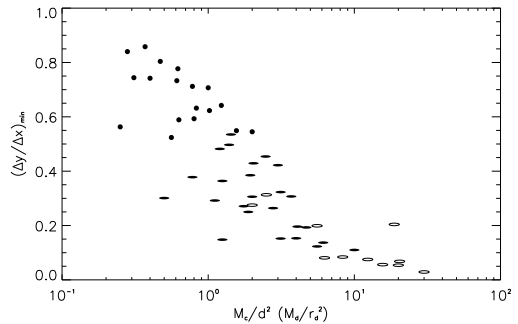}
	\caption{Minimum axis ratio $(\Delta y/\Delta x)$ as a function of initial disk-point mass configuration in terms of $M_c/d^2$  (in units of $M_d/r_d^2$) for the Maclaurin disk simulations.  Plotting symbols correspond to behavior of disk-point mass system as in Figure \ref{fig:md2behavior}.}
	\label{fig:yxmin}
\end{figure}

A final way of representing the difference between the evolution of disks in the presence of a nearby massive particle is to compare the amount of disk distortion which occurs as the disk collapses.  Because the point mass is located on the $x$-axis, tidal forces cause the disks to collapse along the $y$ direction faster than along the $x$ direction.  Therefore, as the disk evolves the ratio of the disks' extent along the $y$-axis to its extent along the $x$-axis decreases from the initial value of $1$. Figure \ref{fig:yxmin} displays the minimum axis ratio $(\Delta y/\Delta x)_{min}$ that each simulated Maclaurin disk acquires during collapse.  As is evident in the plot, the larger the initial force value, the more effectively tidal forces squeeze and stretch the disk as it collapses.  The scatter in this plot is due to the geometric considerations discussed in $\S$ \ref{sec:DependMd2} which cause systems with small separation distances to be more severely distorted due to tidal squeezing. Because the outer portions of a Mestel disk are bound less tightly to the center of mass of the disk, Mestel disks develop significantly smaller values of $(\Delta y/\Delta x)_{min}$  than do the Maclaurin disks with the same values of $M_c/d^2$. 

\subsection{Disk Dispersal}

It is important to note that even a disk which is substantially distorted due to the interaction ultimately forms a dense structure.  For the mass ratios and initial separations considered in this study, the tidal forces are not violent enough to disperse the disk mass into the background GMC. Instead the tidal forces determine the resulting geometry of the dense structure.  Even the most violent interactions between a disk and a point mass result in the disk accreting onto the massive particle. We assume this point mass represents a high density region.  It is likely that this region is gravitationally unstable, or becomes so with the additional accreted mass, and therefore will collapse into a dense structure as well. Thus, for the considered range of initial conditions, gravitational interaction alone is incapable of making dense structures (i.e., the disks) significantly less dense.  The most destructive an interaction can be is to collect nearby gas onto another dense structure.  This effectively moves the dense structures around but does not destroy them.  Consequently, under the influence of gravity alone, dense structures are destined to remain dense structures.   

Our simulations assume that a protocluster starts in an initially cold state, that is there are no internal motions.  Therefore, the initial conditions force our protoclusters to be gravitationally bound objects.  Thus, the final states of the tidally distorted protoclusters are also gravitationally bound.  In order to disperse the mass contained within these dense objects a mechanism other than gravity must be invoked, for example winds from young massive stars formed within or very near to the protocluster.


\subsection{Moving Interactions}

Observations indicate that GMCs are not quiescent static objects but rather they exhibit moderate amounts of internal motion. For GMCs with sizes of $10-50$ pc, the corresponding clump to clump velocities are on the order of a few kilometers per second.  Therefore, in addition to the simulations described above, we complete simulations of the collapse of Maclaurin disks that are initially moving with respect to the nearby massive particle.  

This ensemble of simulations considers systems with particle-to-disk mass ratios of $10-1000$, initial separation distances of $5-10$ pc, and initial relative velocities of $1-10$ km s$^{-1}$.  The angle between the x-axis and disk's velocity vector varies from $15$ to $45$ degrees. We found that for these initial conditions, the disk-point mass systems behaved in much the same way as the corresponding zero velocity systems.  The collapse times for the disks are short ($0.342$ Myr) so that the disks do not travel a significant distance before collapse, and therefore the non-zero velocities do not significantly change the behavior of the disk-point mass system.

For the $60$ simulations completed, over $80\%$ of the systems with initial non-zero relative velocities exhibited behavior patterns in the same category (as defined in $\S$ \ref{sec:IntOutcomes}) as the corresponding system with an initial relative velocity of zero.  Systems that did not exhibit the same behavior as their zero velocity counterparts were slightly more interactive.  Specifically, the system with $M_c = 1000 M_d$, $d = 7 r_d$, and $v_0 > 0$ km s$^{-1}$ accreted onto the nearby point mass.  The disk had a positive initial velocity component in the direction of the point mass.  This effectively reduced the separation distance $d$ and shifted the position of the system in the behavior diagram (Figure \ref{fig:md2behavior}) to the left.


\section{Discussion}\label{sec:Discussion}


\subsection{Initial Central Concentration and Tidal Disruption}\label{sec:DiscussCompare}

The results of our simulations imply that central concentration has a modest effect on the evolution of disks in the presence of a massive perturber.  Specifically the Mestel disks, which are strongly centrally concentrated, have outer regions which are more easily disrupted by the point mass than the Maclaurin disks.  For a given disk to point mass ratio, Mestel disks exhibit accretion onto the point mass and stronger elongation at larger distances than the Maclaurin disks, due in large part to the longer free fall time of the outer portions of the Mestel disk.

While Mestel disks have exteriors that are more easily disrupted, their interiors collapse rapidly producing a dense core ($\S$ \ref{sec:DenseEnhanc}).  In Mestel disks that are tidally stretched into filaments, the result is a very dense inner core with little elongation that is surrounded by an elongated lower density halo of material.  On the other hand, Maclaurin disks which are tidally stretched produce structures that are less centrally peaked and less strongly elongated.  Assuming stellar cluster structure reflects the initial protocluster structure, the ONC is a good example of a dense centrally concentrated cluster with a relatively lower density star forming region surrounding it \citep{ODell2001ARAA}.  NGC 1333 is a good example of a cluster which shows moderate elongation (axis ratio $\sim 2:1$) but a roughly constant density in the central regions. \citep{Gutermuth2008ApJ}.  If NGC 1333 formed from a flattened protocluster disk, it is likely that the disk was less centrally condensed and more closely resembled a Maclaurin disk than a Mestel disk.

An important point to note is that the tidal zone outside of which disks can be considered to collapse as if in isolation is the same for both Maclaurin and Mestel disks.  The bottom line separating the collapse and weak evolution scenarios in the $M-d$ plot (Figure \ref{fig:md2behavior} is the same for both the Maclaurin disks (top panel) and the Mestel disks (bottom panel).  Thus, on the largest scales, disruption of the protocluster disk is not sensitive to the specific surface density distribution.  However, once a protocluster is within the tidal zone, the amount of distortion the disk will experience does depend on the amount of central concentration.  The more centrally condensed disks have exteriors that are more easily disrupted and interiors that are more robust against tidal distortion. 


\subsection{Effects of Other Physical Processes}\label{sec:DiscussOtherProcess}

The effects of pressure have not been included in the $N$-body simulations and will produce some modifications to the results presented here.  As disk collapse proceeds, pressure will increase and at the latest states of collapse may prevent the entire disk from collapsing to a concentrated point.  Therefore, actual protocluster disks will probably not have mean surface densities that are as high as those reached by our simulated disks.  In addition, tidal stretching of the protocluster gas will produce pressure enhancements along the $x$-axis which may results in less compression in the $y$ direction, producing clusters with elongations less extreme than those quoted in $\S$ \ref{sec:Elongation}.

Small deviations from a perfectly smooth initial density distribution will be magnified during collapse and will likely cause fragmentation and lead to star formation \citep{Klessen2000ApJS, Klessen2001ApJ, Heitsch2008ApJ}.  If O and B stars are formed within the protocluster, strong winds produced by these stars may act to disrupt the protocluster from the inside out, possibly halting star formation near the O and B stars and altering the shape of the embeddeding gas. However, these effects take place at a later time in cluster evolution than is considered in this paper.

Our simulations considered protocluster disks composed of particles with zero initial velocity.  However, observations of protocluster sized cloud cores reveal unstructured internal velocities of $\sim 1$ km s$^{-1}$.  These velocities are less than the escape speed from the edge of a $10^3 \msun$ disk with a $1$ pc radius which is $\sim 3$ km  s$^{-1}$, and below the average speed required for the protocluster disk to be in virial equilibrium $\sim 2$ km  s$^{-1}$.  Both the escape speed and the virial speed increase as $R_d^{-1/2}$ as the disk collapses (and $R_d$ decreases). Collapse proceeds rapidly ($\lesssim 0.5$ Myr) once a protocluster disk becomes gravitationally unstable and thus it is unlikely that a significant fraction of the protocluster gas parcels will 'evaporate' due to internal motions. 

These internal motions however, will contribute to the internal structuring of the collapsing cloud.  The angular momentum associated with the internal motions will act to rearrange the density enhancements within the collapsing protocluster and thus perhaps give structure to the forming stellar cluster.  This turbulence may also work to enhance or inhibit star formation within regions of the collapsing disk.  However, unless the internal motions are well ordered so that the total angular momentum of the disk is much greater than zero the angular momentum from the internal motions will not be capable of supporting the protocluster against gravitational collapse, or changing the gross geometry of the cluster.  


\subsection{Applications in Cluster Forming Regions}

With recent advances in infrared telescopes, and most recently, with the Spitzer Space Telescope there has been a shift in our understanding of the distribution of young stars within molecular clouds.  These higher sensitivity observations reveal that clusters of young stars are not usually isolated within molecular clouds but often have neighboring smaller clusters which were previously unidentified.   Examples of regions where clusters appear to be forming near other clusters include the Perseus molecular cloud, Orion A, and the region surrounding Mon R2 (\citealt{Kirk2006ApJ};$~$\citealt{Allen1995PhD}; Gutermuth et al., in prep.) 

Our simulations allow us to estimate the regions around massive clusters where we should not expect smaller clusters to successfully form from flattened protocluster clouds. If we consider the system from the perspective of the more massive disrupting point mass, our calculations allow us to define a ``tidal zone'' around the dense region or massive protocluster that the point mass represents. Within this tidal zone smaller clusters cannot form without significant disruption or distortion.  The tidal zone's size depends on the mass and radius of the collapsing protocluster.  Specifically, the size of the tidal zone around a given massive object increases as the mass of the collapsing protocluster decreases.  In addition, the size of the tidal zone increases as the initial radius of the protocluster increases.  This is consistent with the Roche calculation where $d \sim M_d^{-1/3}r_d$.

Recall that the high density disrupting object is treated as point mass in our simulations.  This overestimates the strength and focusing of the force the disrupting object has on the collapsing protocluster (assuming that the disrupting object is not exactly spherically symmetric).  Therefore, the estimates of the outer region of the tidal zones presented here are upper limits.  That is, the tidal zones may in fact be smaller and the tidal distortion less apparent.

Consider a point mass of $10^4 \msun$ a distance $d$ parsecs away from the center of a $10^3 \msun$ Mestel disk truncated at a radius of $1$ pc.  Figure \ref{fig:md2behavior}  indicates that if the disk center lies within $\sim 2$ pc of the point mass, it will accrete onto the point mass.  However, if the disk center is between $\sim 2$ and $\sim 4$ pc from the point mass, the disk will collapse while being elongated into a filament-like structure.  If the disk center is more than $\sim 4$ pc from the point mass, the disk will collapse axisymmetrically, as if in isolation.  Therefore our calculations indicate that a $10^4 \msun$ cluster should not have surviving neighbor clusters of $10^3 \msun$ within $\sim 2$ pc.  

The Roche limit for a $10^3 \msun$ Mestel disk with a radius of $1$ pc in the presence of a $10^4 \msun$ point mass is $d_{R} = 2.71$pc.  Therefore the Roche limit falls neatly inside the region where disk disruption is significant.  However, our simulated disks exhibit some distortion at slightly larger distances ($\sim 4$ pc) than the analytic calculation.  This is because the Roche calculation ignores the effect of the disk accelerating toward the massive particle as it collapses, and thus underestimates the effect of the gravitational force on the protocluster.

The Maclaurin disk is even more robust against accretion onto the nearby point mass due to its shorter free fall time.  A $10^3 \msun$ Maclaurin disk with radius $1$ pc will accrete onto a nearby $10^4 \msun$ point mass only if the separation distance is less than $\sim 1.5$ pc.  If the separation distance is between $\sim 1.5$ pc and $\sim 4$ pc the Maclaurin disk will collapse into a filament-like structure, whereas if the separation distance is greater than $\sim 4$ pc, the disk will collapse as if in isolation.  Thus, for both the Maclaurin and Mestel disks, the ``tidal zone'' in which collapsing protoclusters of $10^3 \msun$ are compromised is $\sim 4$ pc from a $10^4 \msun$ high density region.

Estimates of this kind are applicable when we consider the distribution of clusters within molecular clouds.  For instance, these estimates are consistent with observations of the Perseus molecular cloud which reveal the cluster B5 forming in close proximity to the IC 348 region.  \cite{Kirk2006ApJ} presented an extinction map of the Perseus star forming region derived from the Two Micron All Sky Survey images created as a part of the COMPLETE survey \citep{Ridge2006AJ}. They identified 11 ``super cores'' within this map, including the B5 and IC 348 star forming regions.  These large, high density cores are similar to the types of protoclusters our simulations considered.

In this region a $10^3 \msun$ cluster (B5) appears to be forming near the $2 \times 10^3 \msun$ IC 348 cluster.  According to Figure \ref{fig:md2behavior}, we should expect to find a $10^3 \msun$ cluster surviving near a $2 \times 10^3 \msun$ only if it is at least $2.5 - 3$ pc away.  Assuming a distance to Perseus of $250 \pm 50$ pc, the separation of IC 348 and B5 is approximately $4.5 \pm 0.9$ pc away (in projection) and therefore B5 is safely outside of the tidal zone surrounding IC 348.

Another region where our estimates are applicable is in the region containing Mon R2.  Observations of this region indicate smaller clusters surrounding the main Mon R2 cluster.  The main Mon R2 cluster contains a total mass (in stars and gas) of $\sim 1700 \msun$ and has an effective rations of $3$ pc (assuming an average stellar mass of $0.5 \msun$ and an extinction to mass conversion of $1$ A$_V$ pc$^{-2} = 15 \msun$).  Therefore, from Figure \ref{fig:md2behavior} clusters with masses greater than $10^3 \msun$ should not form within about $1-2$ pc of the main Mon R2 cluster.   No clusters are found within $1-2$ pc of the main Mon R2 cluster.  In fact, the nearest region forming groups of stars is $\sim 3.7$ pc (in projection) to the east of the center of Mon R2 and has a mass of $\sim 135 \msun$ and a radius of $~1$ pc (Gutermuth et al., in prep).  Using the Roche limit calculation, we find that the tidal zone in which a $135 \msun$ cluster with radius may be disrupted in the presence of a $1700 \msun$ point mass is $2 - 3$ pc depending on the disk density distribution chosen.  Assuming the results of our simulations scale down in mass as the Roche calculation does, the estimated tidal zone is somewhat larger, $\sim 4$ pc.  Thus we predict that a protocluster at a distance of $3.7$ pc should collapse as if in isolation and not be significantly disrupted by the neighboring $1700 \msun$ cluster.

In regions containing significant amounts of distributed gas, such as the region in L1641 south of the Orion Nebula Cluster (ONC), collapse disruption due to nearby massive objects (in this case the ONC) may be less effective than estimated by these simulations.  In this region there is a string of stellar clusters that have developed near one another and near the ONC \citep{Allen1995PhD, Allen2007PPV}.  Dense gas distributed around the protocluster is likely to wash out the gravitational force felt by a collapsing protocluster due to a nearby massive object.  This may allow clusters to form nearer each other than one would estimate solely based on the mass and proximity of the nearest massive neighbor.  Therefore, care should be taken when applying the estimates presented here to significantly gas-rich regions.

A rough estimate of the average background density required to change the results of our simulations can be made by assuming the background gas has a uniform density $n_0$. The force on a protocluster from a $10^4 \msun$ point mass a distance $4$ pc from the protocluster center equals the force on the protocluster due to sphere of gas with radius $4$ pc centered $4$ pc from the protocluster if $n_0 \approx 750$ cm$^{-3}$ (assuming the gas is composed of molecular hydrogen).  $4$ pc is the edge of the tidal zone for a $10^3 \msun$ protocluster near a $10^4 \msun$ point mass.  Thus background gas densities higher than this may begin to wash out the effects of nearby massive perturbers.  However, the densities required for the forces to be equal scale as $n_0 \sim d^{-3}$.  So for the background gas to have as much influence on the disk as a $10^4 \msun$ point mass $2$ pc from the protocluster, a much higher average gas density of $n_0 \approx 6000$ cm$^{-3}$ is required.  Typical densities in GMCs range from $10^2$ to $10^3$ cm$^{-3}$ \citep{Liszt1981ApJ}. In the higher density  regions, which often exhibit clustered star formation, densities may reach $5-15 \times 10^3$ cm$^{-3}$ \citep{Kirk2006ApJ}, and so it is in these denser regions of GMCs that our simulations may overestimate the effects of nearby perturbers on protocluster evolution.

Tidal zone estimates of the kind presented here may be applied to other rich star forming regions such as the Cyguns X region which contains many clusters as well as many sites of current star formation \citep[][and references therein]{Schneider2006AA}. 


\section{Summary}

Our calculations of the collapse of protoclusters in GMCs provide the framework in which to identify regions where protocluster collapse can be assumed to proceed as if in isolation.  The following conclusions can be drawn from the results of our simulations:
\begin{enumerate}
\item Depending on the geometry and strength of the interaction, protocluster disks either collapse as if in isolation, are weakly or strongly elongated, or are accreted by the neighboring point mass.  The outcomes of our simulations are presented in Figure \ref{fig:md2behavior} and are consistent with the Roche limit estimates of disruption.  
\item The amount of disk disruption due to a nearby massive particle depends most sensitively on the mass of and distance to the particle and less sensitively on the distribution of material within the protocluster disk.  
\item The tidal zone defined by the separation between \emph{collapse} and \emph{elongation} evolution scenarios does not sensitively depend on how centrally concentrated the disk is.  Maclaurin (weakly centrally concentrated) and Mestel (strongly centrally concentrated) disks have tidal zones that are roughly the same for a given disk to point mass ratio.
\item The interior of disks that are initially more centrally condensed are more robust against tidal distortion than those that are less centrally condensed.  Therefore, for flattened protoclusters that formed from density distributions that were centrally peaked, it is likely that once the it becomes unstable against gravitational collapse, the inner portion of the disk will continue to collapse, even in the presence of another massive object.  
\item On account of both the short collapse times of protoclusters and the modest clump to clump velocities observed in GMCs, the interaction outcomes do not depend significantly on the relative velocity between the protocluster and the point mass.  
\item In systems with conditions consistent with those observed in GMCs, gravitational interactions alone only produce density enhancements within the protoclusters.  Thus, tidal stripping does not significantly disperse mass from the condensing disk.
\end{enumerate}


The authors thank F. Adams, L. Allen, and R. Gutermuth for many beneficial discussions, and the anonymous referee for a careful and constructive review of the manuscript.   The bulk of this work was during a Visiting Student appointment at the Harvard-Smithsonian Center for Astrophysics.  E. Proszkow is grateful for the hospitality of  L. Allen who made the visit possible.  This work was supported by the University of Michigan through the Michigan Center for Theoretical Physics, and by the Spitzer Space Telescope Theoretical Research Program (1290776).  Additional support for this work, part of the Spitzer Space Telescope Theoretical Research Program, was provided by NASA through a contract issued by the Jet Propulsion Laboratory, California Institute of Technology under a contract with NASA (1288820).  Support for P. Myers was provided by the NASA grant ``Motions and Initial Conditions in Star-Forming Dense Cores'' (613449, NAG5-13050).  


\section{Appendix}

In this Appendix we discuss the use of $N$-body simulations to calculate the dynamics of a collapsing gaseous disk.  We provide the results from the simulated collapse of isolated Maclaurin and Mestel disks using an $N$-body code.  Analytic treatments of collapsing gaseous systems provide generalized descriptions of collapse behavior by employing idealized geometries, density distributions, simplifying approximations, etc.  However, analyses of this kind are limited by the very approximations that make them possible.  On the other hand, fluid dynamics codes are capable of simulating various initial geometries and may include more complex physical processes such as gas pressure, drag forces, or interactions with magnetic fields, etc. $N$-body simulations are computationally much more efficient than fluid dynamics codes and can be used with care to simulate the behavior of gaseous (or gas dominated) systems. \cite{Toomre1972ApJ} is a classic example of using an $N$-body simulation (specifically, a restricted three-body simulation) to study the dynamics of interacting galaxies.  More recently, full $N$-body simulations have been employed to study galaxy interactions in the stellar-dynamical limit \citep{Hernquist1993ApJ}.

Modeling the behavior of a gaseous disk with an $N$-body code requires that care be taken when configuring the initial conditions and interpreting the results of the simulation.  The initial configuration should minimize the scattering interactions that occur.  This is accomplished by choosing an axisymmetric ring configuration where particles in each ring have equal mass. Specifically, the simulated disk contains a central point mass and 25 rings of 32 equal mass particles.  The total mass of each ring is then varied to produce a a $10^3 \msun$ disk with radius $1$ pc a desired initial density distribution.

The number of particles $N$ chosen to represent the protocluster was determined by two competing factors: $N$ must be large enough that the particles represent the smoothness of the density distribution and small enough to minimize particle-particle scattering interactions.  We found that a choice of $N \sim 800$ produces a disk distribution that was both smooth enough to model accurately the behavior of the isolated disks and sparse enough to minimize the number of scattering interactions which occurred during the collapse simulation. For example, a disk with $N \sim 500$ particles had initial mass and force profiles with errors larger than the $N \sim 800$ disk.  We compared simulations of disks with $N \sim 2000$ and $N \sim 800$ particles.  We found that the evolution of both disks were very similar during the first half of the collapse, but at later times the $N \sim 2000$ disk had larger errors in their mass profiles and collapse times (when compared to the analytic solution) due to more frequent scattering interactions in the higher number density regions of the disk.

Another consideration when calculating the dynamics of fluids with $N$-body simulations is that the simulated behavior and conclusions drawn from that behavior should be consistent with the behavior of fluid systems.  A consequence of using the $N$-body simulation is that as the disk of particles collapses, particle-particle interactions will allow some particles to pass through the center of mass and continue on a radial path outward.  This collapse and subsequent expansion is an artifact of the simulation method.  In a fluid system, mass collapses directly toward the center of mass until the pressure gradient becomes too large and collapse is slowed.  Therefore, to simulate accretion in the $N$-body code, we have implemented a sink cell algorithm.  

This algorithm allows the central particle in the disk to inelastically ``absorb'' the mass of any particle that comes within the sphere of radius $r_{sink}$ surrounding the central particle.  Experimentation with values of this fiducial radius resulted in the choice of $r_{sink} = 0.049$ pc $\approx 10,000$ AU.  We required that the sink cell radius be small enough that it encompasses only the central particle at $t = 0$, and large enough that it maximized the number of disk particles that entered the sink cell during the simulation.  For the test calculations of isolated Maclaurin and Mestel disks, more than $90\%$ of the disk particles collapse from their initial position within the disk directly into the sink cell.

We model the continuous disk as a set of discrete particles arranged in an axisymmetric ring structure.  The ring spacing, ring mass, and central particle mass are chosen so that these disks have radially averaged mass and force profiles that are in close agreement with a Maclaurin and Mestel profiles. Specifically, the simulated Maclaurin and Mestel disks have errors of $2.4\%$ and $3.2\%$ in their initial mass profiles, respectively.  The initial force profiles have errors of $6.0\%$ and $10.0\%$, respectively.  These percent errors in the mass and force profiles are calculated as the r.m.s. error in the profile divided by $M(r_{1/2})$ and $F(r_{1/2})$ where $r_{1/2}$ is the initial half-mass radius of the Maclaurin or Mestel disk.

\begin{figure}
  \plotone{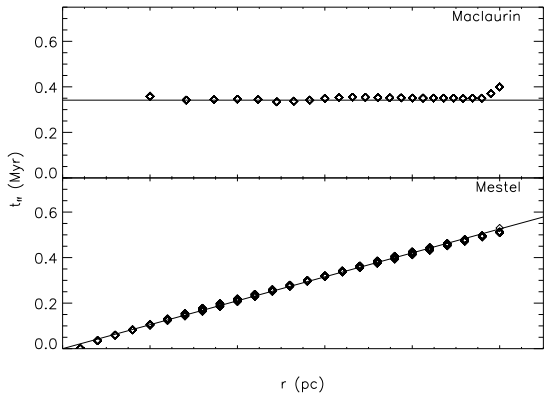}
	\caption{The free fall time for each particle in the simulated collapsing disk is plotted as a function of the initial radial position of the particle within the disk (diamonds).  The upper (lower) panel displays the data from the Maclaurin (Mestel) disk simulation.  Solid lines correspond to the analytic free fall time calculated from the collapse solution for each initial surface density profile.}
	\label{fig:tff2}
\end{figure}

\begin{figure}
  \plotone{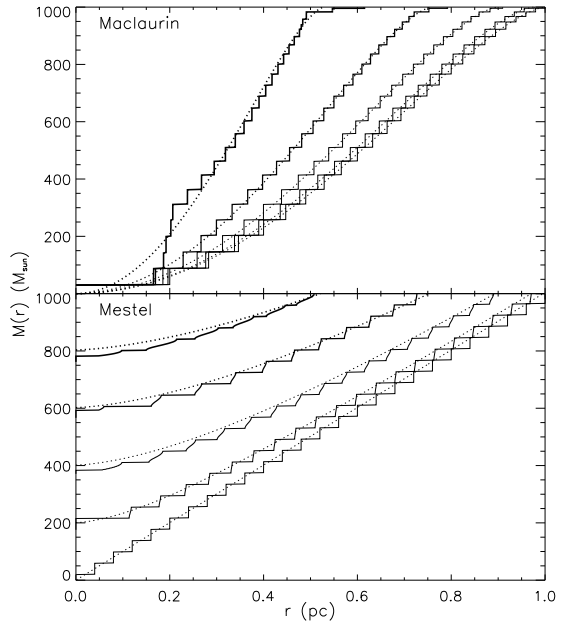}
	\caption{Top panel displays cumulative mass distributions for the Maclaurin disk at times $t = 0, 0.2, 0.4, 0.6$, and $0.8 ~t_{ff}$ (in order from right to left).  Thicker curves correspond to later times.  The solid curve corresponds to the simulated data while the dashed curve corresponds to the analytic solution for the cumulative mass distribution at the corresponding times.  Cumulative mass distributions for the Mestel disk at the same times are displayed in the bottom panel.}
	\label{fig:massprof}
\end{figure}

One of the problems inherent in representing a fluid disks as set of discrete particles is that the inner and outermost portions of the particle disk suffer from ``edge effects'' caused by the discrete nature of the ring structure.  That is, these portions of the disks have mass and force profiles that differ from the edge of the analytic solutions.  These ``edge effects'' are visible in the initial mass profiles depicted in Figure \ref{fig:massprof} and in the calculated free fall times in Figure \ref{fig:tff2}.  We found that these effects do not propagate throughout the system but remain confined to their respective regions of the disk.  Therefore, we can overcome these ``edge effects'' by describing the evolution of the disk without considering the inner and outermost regions in the description.  

Thus, throughout this paper the descriptions of disk behavior do not include the innermost or outermost ring of the Maclaurin disk.  The region of interest for collapse behavior is $r_{1} < r < r_{25}$ for a disk with 25 rings, where $r_i$ is the radius of the $i$th ring.  The Mestel disk has the property that the equation of motion for a particle in the disk depends only on the mass interior to the ring containing that particle \citep{Mestel1963MNRAS}.  This is not generally true for disk systems, but the Mestel disk is a special case where Gauss's law in disk geometry does hold.  This allows us to embed the Mestel disk we are interested in simulating within a larger Mestel disk without changing the behavior of the smaller interior disk.  Therefore, instead of ignoring the outer rings in the determination of disk behavior, we initially embed the Mestel disk in a larger disk with $R_{large} = 1.5r_{Mestel}$, which effectively moves the ``edge effects'' outside of the region of interest ($r \leq r_{Mestel}$), and then consider the entire small Mestel disk except for the innermost ring ($r_{1} < r \leq r_{Mestel}$).  It is worth noting that more than $90\%$ of the mass within the Maclaurin and Mestel disks is contained within the regions of interest defined above and thus only a small fraction of the disk suffers from these ``edge effects''.

The test calculations of disk collapse considered isolated Mestel and Maclaurin disks with initial radii of $r_d = 1.0$ pc and total masses of $M_d = 1000 \msun$. Analytic solutions to the pressure-free collapse of these systems exist and we compare our simulated results to the analytic solutions.  For these initial conditions, the free fall time for a Maclaurin disk $0.342$ Myr independent of initial position within the disk.  The Mestel disk has a free fall time for a test mass initially at $r$ of $t_{ff}(r) = 0.526 r$ Myr.  The resulting free fall times for the simulated disks are plotted in Figure \ref{fig:tff2} as a function of initial position of the particle within the disk.   The free fall time for a particle is defined as the time that the particle enters the sink cell.  For particles that do not enter the sink cell, the free fall time is the time at which the particle passes closest to the disk's center.

The ``edge effects'' are clearly seen in the free fall times of the inner and outer rings in the Maclaurin disk.  The error in the calculated free fall times is $3.4\%$ and $2.4\%$ for the Maclaurin and Mestel disks respectively.  This error is calculated by dividing the r.m.s. error by the free fall time of a particle initially at the half-mass radius.  By choosing to consider the regions of interest described above, the errors in the simulated behavior are reduced.  For instance, if the entire disk was considered in the calculation of the free fall times, the errors for the Maclaurin and Mestel disks would be $4.6\%$ and $2.7\%$, respectively.

Figure \ref{fig:massprof} depicts the evolution of the cumulative mass profiles for the collapsing Maclaurin (top panel) and the Mestel (bottom panel) disks.  The solid curves correspond to the cumulative mass profiles of the simulated disk at $t = 0.0, 0.2, 0.4, 0.6$, and $0.8 t_{ff}$.  The dashed curves correspond to the analytic solution for the cumulative mass profiles of the collapsing disk at the corresponding times.  It is clear from this plot that discrepancies in the mass profile which are relics of the initial discrete particle configuration remain throughout the simulation. However, the error is contained in the outermost rings, and does not significantly affect the evolution of the bulk of the disk.  At each time step, the mass profile agrees well with the analytic solution out to a fraction of the disk radius, and this fraction does not change significantly over time.  The errors in the mass profiles range from $2.4\% - 6.1\%$ and $3.2\% - 7.4\%$ for the Maclaurin and Mestel collapse simulations respectively for $t \leq 0.8 ~t_{ff}$, where the percent error in the mass profile is defined as the r.m.s. error in the profile divided by half of the total disk mass. 

The error in particle positions was also calculated as a function of time.  At a particular time and for a particular ring of particles in the disk, we calculate the r.m.s. error in the positions of the particles with respect to the analytic collapse solution.  The r.m.s. error is scaled by the distance that the ring has collapsed, resulting in a percent error in the position of a ring at a particular time.  For the Maclaurin disk, the errors in the simulated ring positions range from $0.13-39.19\%$ (with a mean of $7.04\%$) if we consider all rings, and all times.  However, the largest percent errors occurred in the inner-most rings and at times close to the collapse time for the ring.  If instead, we consider the region of the disk between rings 3 and 23 and times less than $0.75 t_{ff}$ the errors in ring positions range from $0.13\% - 9.95\%$ (with a mean of $5.01\%$).  This calculation was also completed for the simulated Mestel disk.  Considering all rings and times, the errors in the simulated particle positions ranged from $0.14-16.88\%$ (with a mean of $5.53\%$).  If only the portion of the disk between rings 3 and 24, and times less than $0.75 t_{ff}$ are considered, the errors in ring positions range from $0.14-12.46\%$ (with a mean of $5.11\%$).  Therefore, the bulk of the particles in the simulated Maclaurin and Mestel disks have small positional errors throughout most of the simulation.

\begin{deluxetable}{lcc}
\tablewidth{0pt}
\tablecaption{Simulation Errors\label{tab:Errors}}
\tablehead{
\colhead{ } & \colhead{Maclaurin Disk} & \colhead{Mestel Disk}} 
\startdata
Initial Force Profile, $F(r)/m$                  & 6.0\%       & 10.\%       \\
Free Fall Time, $t_{ff} $                 & 3.4\%       & 2.4\%       \\
Initial Mass Profile, $M(r,t=0)$                & 2.4\%       & 3.2\%       \\
Mass Profiles, $M(r,t \leq 0.8 t_{ff})$  & 2.4 - 6.1\% & 3.2 - 7.4\%
\enddata
\end{deluxetable}

These test simulations provide evidence that our $N$-body code is indeed useful for studying systems of collapsing disks.  We were able to reconstruct with our simulations disks of discrete particles that behaved as the corresponding gaseous disk with the smoothed surface density distribution behaves.  These simulations brought to light the ``edge effects'' that our discrete systems displayed while at the same time revealing that those effects truly remained on the edge of the disks. Therefore, they can be consciously worked around by defining the region of interest as the inner portion of the disk.  In addition, the particles in our simulation underwent very few particle encounters during the time leading up to the disk collapse indicating that the choice of the axisymmetric ring distribution succeeded in eliminating the scattering effect of random particle-particle interactions. And finally, our calculated initial force profiles, free fall times, mass profiles, and particle positions agreed well with analytic solutions justifying use of this simulation method for calculation of Maclaurin and Mestel disk systems.


\end{document}